\documentstyle[twocolumn,prb,aps,psfig]{revtex}
\begin{document}

\title{Analytical Long-Range Embedded-Atom Potentials}
\author{Qian Xie} 
\address{Max-Planck-Institut f\"{u}r Physik komplexer Systeme, Bayreutherstrasse 40,
Dresden, D-01187, Germany}
\author{Wen-qing Zhang and Nan-xian Chen}
\address{Institute of Applied Physics, Beijing University of Science and
Technology, Beijing, 100083, China} 
\twocolumn[
\maketitle
\begin{quote}
\parbox{16cm}{\small
We present a systematical method for obtaining analytical long-range 
embedded-atom potentials based on the lattice-inversion method.
The potentials converge faster (exponentially) than Sutton and Chen's  
power-law potentials (Philos. Mag. Lett. {\bf 61}, 2480(1990)). 
An interesting relationship between the embedded-atom 
method and the universal binding energy equation of Rose et al. 
(Phys. Rev. B {\bf 29}, 2963 (1984)) is also pointed out. 
The potentials are tested by calculating the elastic constants, 
phonon dispersions, phase stabilities, surface properties and melting
temperatures of the fcc transition metals. The results are overall in 
agreement with experimental or available ab initio data.\vspace{0.3cm}

}
\end{quote}]

\section{Introduction}

\begin{figure}
\centerline{\psfig{figure=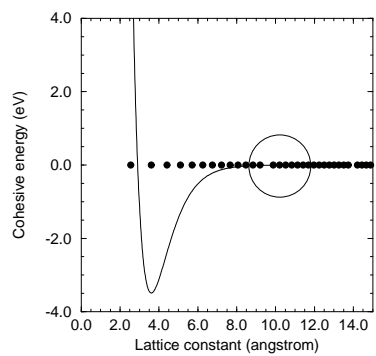,height=6cm,angle=0}}
\caption{
Schematical illustration for the need of long-range potential. The
filled cirlces denote the radial distribution of the atom shells of fcc
copper. The curve is the UBER for copper. The big
circle shows the region to which the potential range should be extended.}
\label{fig:why need}
\end{figure}

There are quite a few problems in atomistic simulation for which
long-range potentials are needed. An important one is the problem of 
structural energy difference (SED). 
Normally the minimum SED is of the magnitude of one percent of the 
cohesive energy or so. Evidently if the potential range is
only up to the second nearest neighbors, then a pair functional
model will predict no energy difference between the fcc and hcp structures. 
We have to extend the ranges of potentials to further distance.
Usually people impose a cut-off on the potentials and
adjust the model parameters so that correct SED can be produced.
However, the unphysical cut-off procedure thus becomes the dominant factor
for predicting the SED: Suppose we fix the model parameters
and change the cut-off distance, then it is highly possible to
find that the SED varies in sign with respect to the cut-off distance(see
Fig.1 in Ref. 1). The safe way to remove this drawback is to extend the 
range of the potentials so that the contributions from the furthest atoms
become less than one percent of the cohesive energy. 
Fig. \ref{fig:why need} illustrates that 
to get reliable SED between fcc and bcc for copper the potential range should 
be extended into the big circle (corresponding to some tolerable error bound). 
Only to that region
(and beyond) does the universal binding energy relation (UBER) of Rose et al.
~\cite{rose-prb} decrease to the magnitude of the SED between fcc 
and bcc lattices. For alloys the problem will be more complex. There are some
superstructures with very large unit cells. To calculate the heats of
formation for these competing structures needs very long-range potentials.
For calculating the elastic constants, the potential range is
also important. For example, the predicted shear modulus $C'$ of bcc
structure is zero if a nearest-neighbor potential is used, so
a potential range beyond nearest-neighbor distance is required for
bcc structure. Long-range potentials are also needed for phonon
calculation. In the case that the unit cell is very large, the
potential range should be long enough so that all the atoms
in the unit cell can interact with each other hence the force constants
linking them do not vanish.

To determine interatomic potentials, one has to assume some functional 
forms for them (such as exponential and power-law functions), and the 
potential parameters are fitted from experimental properties.
The parameters can be exactly solved within Johnson's nearest-neighbor model
with exponential potentials~\cite{johnson-prb1} and Sutton and Chen's 
long-range model with inverse power-law potentials~\cite{sutton}. Of course,
Johnson's model is not applicable in many cases because it is too
short-range. On the other hand, as inverse power-laws, Sutton and Chen's 
potentials converge, however, slower than the exponentials (this will be
explained later). In molecular dynamics simulation, 
slow convergence of the potentials will result in the increasing of the 
time needed to make the neighbor list. 
In a long-range model that does not employ power-law functions, 
the conventional way to obtain the parameters is the numerical method 
of least-square fitting. The disadvantage of the numerical fitting method
is that it is a little arbitary. Different authors may obtain different
parameters, or the same author may obtain different parameters at 
different time, while a small variation of the parameters may lead
to change of the long-range tail remarkable for the SED. 
This makes the conventional fitting method problematical when the 
fitted potentials are to be used to calculate something like phase 
stability and stacking fault energy. These properties are quite 
sensitive to the long-range tail which is,
however,  not very refined in the conventional fitting method. 

In this paper, we present a systematical method of obtaining analytical
long-range potentials with satisfactory convergence based on the
lattice-inversion method (LIM). The LIM was first used by
Carlsson, Gelatt and Ehrenreich (CGE) to get parameter-free 
pairwise potential from ab initio total energy calculation.~\cite{carlsson-pma} 
Recently, the inversion formula of CGE was recasted into a concise formulism
by Chen based on the M\"{o}bius inversion transform in number theory
~\cite{chen-prl,chen-pla,chen-pre}. Nevertheless, the original two-body 
inversion scheme has, of course, some problems because of the lack of
many-body contribution. Therefore,
some many-body inversion schemes based on the $N$-body potential~\cite{finnis}
and angularly-dependent Stillinger-Weber potential~\cite{sw} were
developed by Xie and Chen~\cite{xie-prb} and Bazant and Kaxiras
~\cite{kaxiras}, respectively. 

This paper is organized as follows.
In Section II, we briefly introduce the LIM. In Section III, we
present the lattice-inversion model for the embedded-atom method 
(EAM)~\cite{daw-prl}. 
In Section IV, we discuss the parametrization procedure in detail. 
In Section V, we present some calculated results. 
The paper is concluded in Section VI.

\section{The lattice-inversion method}

The LIM can be traced to an early work by CGE.~\cite{carlsson-pma} The
idea is to invert a function from its lattice sum which is
sometimes easier to be obtained. For example,
in the pair potential model (PPM), the cohesive energy can be written
as the summation of the pair potential over the crystal lattice

\begin{equation}
E(R_1)=(1/2)\sum_{m=1}^{\infty}w_mV(p_mR_1)
\label{eq:ppm}
\end{equation}
where $w_m$ is the number of atoms on the $m$-th shell, $p_m$ is the
ratio of the radius of the $m$-th shell to the nearest-neighbor
distance. The cohesive energy as a function of lattice spacing
can be calculated by using first-principles method, or simply taken as
the UBER. Then as CGE suggested, one can use the following inversion
formula to obtain the so-called ab initio pair potential $V(r)$

\[V(r)=\frac{2}{w_1}E\left(\frac{r}{p_1}\right)-\sum_{m=2}^{\infty}
\frac{2}{w_1}\frac{w_m}{2}\frac{2}{w_1}E\left(\frac{p_mr}{p_1^2}\right)\]

\begin{equation}
+\sum_{m,n=2}^{\infty}\frac{2}{w_1}\frac{w_m}{2}\frac{2}{w_1}\frac{w_n}
{w_1}\frac{2}{w_1}E\left(\frac{p_mp_nr}{p_1^3}\right)-
\cdots
\label{eq:cge}
\end{equation}
The multiple summations make the ordering for the inversion coefficients not
obvious. It was not until recently that Chen put forward his elegant
M\"{o}bius inversion formula on three-dimensional crystals.~\cite{chen-pre}
The Chen-M\"{o}bius inversion formula is very simple

\begin{equation}
V(r)=2\sum_{m=1}^{\infty}\mu_mE(p_mr)
\label{eq:inversion}
\end{equation}
The M\"{o}bius coefficients $\mu_m$ can be determined by

\begin{eqnarray}
\mu_1 &=& {1}/{w_1}\nonumber\\
\mu_m &=& -({1}/{w_1})\sum_{p_k|p_m,k\ne m}
\mu_kw_l \hspace{0.2cm}(m\ge 2)
\label{eq:mobius}
\end{eqnarray}
where $l$ is the natural number which satisfies
$p_l=p_m/p_k$. Obviously, Chen's formula requires the set
${\bf P}=\{p_m|m\in{\bf N}\}$ should be a multiplication-close one, i.e.,
given two arbitrary elements $p_i$,$p_j$ $\in$ {\bf P}, their product
$p_ip_j$ should be in {\bf P} too. Actually the
crystal lattices sc, bcc, fcc, hcp and diamond etc. do not 
automatically satisfy this requirement. Therefore, before applying 
the Chen-M\"{o}bius formula
we have to at first construct a close set ${\bf Q}$, which should include
at least part of the elements in the orginal set ${\bf P}$. This task
is easily done for sc and fcc: For them the set ${\bf P}$ is
simply $\{\sqrt{i^2+j^2+k^2}R_1|i,j,k\in {\bf Z},i^2+j^2+k^2\ne 0
\}$, we can construct a
new set ${\bf Q}$=$\{\sqrt{n}R_1|n\in{\bf N}\}$, which covers ${\bf
P}$. The numbers of atoms on the shell $\sqrt{n}R_1$ vanish if
$n$ cannot be written as the square sum of three natural numbers
$i^2+j^2+k^2$. But for other lattices such as bcc, it is difficult
to find a natural close set which covers all the elements in the set ${\bf
P}$= $\{\sqrt{i^2+j^2+k^2}a|i,j,k\in{\bf Z},i^2+j^2+k^2\ne 0\}\cup
\sqrt{(i+1/2)^2+(j+1/2)^2+(k+1/2)^2}a|i,j,k\in{\bf Z}\}$,
where $a$ is the lattice constant. However, for the present physical 
problem we do not have to construct a close set covering all the elements. 
Note that the expansion of eq.(\ref{eq:ppm}) 
should be convergent. That is to say, usually we can truncate at 
some shell,
say, the $M$-th shell, beyond which the function $V(r)$ has
become small enough to be neglected. So we can approximate
eq.(\ref{eq:ppm}) by $E(R_1)=\sum_{m=1}^{M}w_mV(p_mR_1)$, then
we have a set with $M$ elements ${\bf P}$= $\{p_1,p_2,\cdots,p_M\}$.
We can easily generate a close set ${\bf Q}$ which covers ${\bf P}$:
${\bf Q}$= $\{p_1^{k_1}p_2^{k_2}\cdots p_M^{k_M}|k_1,k_2,\cdots,k_M
=0,1,2,3,\cdots,k_1+k_2+\cdots+k_M\ne 0\}$. Re-ordering this close
set from the smallest element to the biggest one, we get the set
as ${\bf Q}$= $\{p_m^{'}|p_m^{'}<p_{m+1}^{'}, m\in {\bf N}\}$. Then
we can rewrite eq.(\ref{eq:ppm}) as $E(R_1)=(1/2)\sum_{n=1}^{\infty}
w_n^{'}V(p_n^{'}R_1)$, where $w_n^{'}=w_m$ when $p_n^{'}=p_m$ and
vanishes when $p_n^{'}$ equals none of the elements in {\bf P}. The
inversion is simply the same as eq.(\ref{eq:inversion}), with the
M\"{o}bius coefficients $\mu_n^{'}$ determined from $w_n^{'}$. 

\begin{figure}
\centerline{\psfig{figure=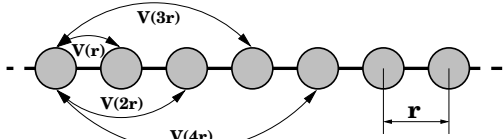,height=2cm,angle=0}}
\caption{A physical mapping of the M\"obius inversion theorem 
in arithmetic number theory.}
\label{fig:mapping}
\end{figure}

In the one-dimensional case, eq.(\ref{eq:inversion}) becomes the
number-theoretic M\"obius inversion formula

\begin{equation}
V(r)=\sum_{n=1}^{\infty}\mu(n)E(nr)
\end{equation}
where $\mu(n)$ is the number-theoretic M\"obius function. This 
physical mapping (Fig.\ref{fig:mapping}) was first dicovered 
by Chen~\cite{chen-prl}.

Very recently, Bazant and Kaxiras have presented a novel scheme to obtain
effective angularly-dependent many-body potentials for covalent 
materials by inversion of cohesive energy curves from many 
configurations~\cite{kaxiras}. Their method, together with our previous 
work for the EAM~\cite{xie-prb}, have shown the potentiality of the LIM 
as a shortcut to obtain the effective interatomic potentials.

\section{The lattice-inversion embedded-atom model}

Generally the physical properties we use to fit the potential
parameters are related more closely to the lattice sums than to
the individual potential function themsleves. Given the
lattice sums of the pair potential $V(r)$ and electron density
$f(r)$

\begin{eqnarray}
\sum_{i}f(R_i)&=&\rho(R_1)\\
\label{eq:rho}
(1/2)\sum_{i}V(R_i)&=&\Phi(R_1)
\label{eq:phi}
\end{eqnarray}
We can easily rewrite the formulas of the bulk modulus and Voigt shear
modulus

\begin{eqnarray}
B=\frac{1}{18\Omega}\{\sum_iR_i^2[V_{\rm eff}^{''}(R_i)&&
-\frac{1}{R_i}V^{'}_{\rm eff}(R_i)]
\nonumber\\
&&+2F^{''}(\rho_e)[\sum_iR_if^{'}(R_i)]^2\}
\label{eq:bulk modulus}\\
G=\frac{1}{30\Omega}\sum_iR_i^2[V_{\rm eff}^{''}(R_i)&&
-\frac{1}{R_i}V^{'}_{\rm eff}(R_i)]
\label{eq:shear modulus}
\end{eqnarray}
as

\begin{eqnarray}
B&&=\frac{1}{9\Omega}\{R_1^2[\Phi^{''}(R_1)+F^{'}(\rho)\rho^{''}(R_1)]
\nonumber\\
&&-R_1[\Phi^{'}(R_1)+F^{'}(\rho)\rho^{'}(R_1)]
+F^{''}(\rho)R_1^2[\rho^{'}(R_1)]^2\}
\label{eq:b1}\\
G&&=\frac{1}{15\Omega}\{R_1^2[\Phi^{''}(R_1)+
F^{'}(\rho)\rho^{''}(R_1)]
\nonumber\\
&&-R_1[\Phi^{'}(R_1)+F^{'}(\rho)\rho^{'}(R_1)]\}
\label{eq:g1}
\end{eqnarray}
where $\Omega$ is the atomic volume, $V_{\rm eff}(r)$ is the effective
pair potential $V_{\rm eff}(r)=V(r)+2F^{'}(\rho)f(r)$. And for transforming
eqs.(\ref{eq:bulk modulus}) and ({\ref{eq:shear modulus}) to the effective
nearest neighbor forms of eqs.(\ref{eq:b1}) and (\ref{eq:g1}),
the following relationship has been used: if
there is a function $h(r)$ whose lattice sum is another function
$H(R_1)$, then

\begin{equation}
\sum_iR_i^nh^{(n)}(R_i)=R_1^nH^{(n)}(R_1)
\end{equation}
Thus we have an equation related to the Cauchy pressure

\begin{equation}
9B\Omega-15G\Omega=F^{''}(\rho)R_1^2[\rho^{'}(R_1)]^2
\label{eq:cauchy}
\end{equation}

On the other hand, the vacancy-formation energy

\begin{equation}
E_v=-\Phi+\sum_i[F(\rho-f(R_i))-F(\rho)]+E_{\rm relax}
\end{equation}
can be approximately written as the lattice sum of the effective
pair potential $E_v=(1/2)\sum_iV_{\rm eff}(R_i)$, since the
numbers of atoms on the shells are much greater than 1, and the
negative relaxation energy further reduces the error. This approximation
has been checked by a simple relaxation calculation in which only the
nearest neighbor atoms around the vacancy site are allowed to relax.
It is found that in the case of copper the calculated unrelaxed
vacancy-formation energy is 1.34 eV, while the relaxed result is
1.31 eV, closer to the experimental value.
Hence, the difference between the vacancy-formation energy and the
sublimation energy can be written as

\begin{equation}
E_s-E_v=F^{'}(\rho)\rho(R_1)-F[\rho(R_1)]
\label{eq:vacancy}
\end{equation}

We can see from eqs.(\ref{eq:cauchy}) and (\ref{eq:vacancy}) that the
nonlinearity of the embedding function reflects the many-body nature
of the embedded-atom potential. If $F(\rho)$ is a linear function with
respect to $\rho$ (corresponding to a PPM), then we have
$3B=5G$ (the Cauchy relation) and $E_s=E_v$, which are the two well-known
drawbacks of the PPM.

The nearest-neighbor distance of the equilibrium lattice can be
obtained by minimizing the total binding energy

\begin{equation}
\Phi^{'}(R_{1e})+F^{'}(\rho_e)\rho^{'}(R_{1e})=0
\end{equation}

Another condition we need to consider is the normalisation for the electron
density. Integrating both sides of eq.(\ref{eq:rho}) with respect to
$R_1$

\begin{eqnarray}
\sum_m\frac{w_m}{p_m^3}\int_0^{\infty}f(p_mR_1)4\pi 
&&(p_mR_1)^2d(p_mR_1)
\nonumber\\
&&=\int_0^{\infty}4\pi R_1^2\rho(R_1)dR_1
\end{eqnarray}
Note that the electron density $f(r)$ should be normalized
$\int_0^{\infty}4\pi r^2f(r)dr=N$ (where $N$ is the number of
electrons), we obtain

\begin{equation}
\int_0^{\infty}4\pi R_1^2\rho(R_1)dR_1=S(3)N
\end{equation}
where $S(3)=\sum_mw_m/p_m^3$. In alloy case, the parameter $N$ should 
be determined by considering the charge transfer. This consideration
is based on empirical Miedema's equation which well decribes the heats
formation of binary alloys. The attractive term in Miedema's equation is
re-interpreted by Pettifor as the contribution of the electronegativity
difference, which is related to the charge transfer.~\cite{pettifor}

The embedding function in the present model is assumed to be a
power-law one

\begin{equation}
F(\rho)=-A\rho^{1/\lambda}
\label{eq:embedding function}
\end{equation}
$\lambda=1$ corresponds to the PPM, while $\lambda=2$
corresponds to the $N$-body potential of Finnis and Sinclair.~\cite{finnis}
In the following section we shall show that with appropriate
functional forms for the electron density and pair potential, this
embedding function will produce exactly the UBER,
and the parameter $\lambda$ is insensitive to the functional
forms of the electron density and pair potential.

The electron density and pair potential in the present model are
structure-dependent as we can see from their inverted formulas

\begin{eqnarray}
f(r)&=&\mu_1\rho(p_1r)+\mu_2\rho(p_2r)+\mu_3\rho(p_3r)+\cdots
\label{eq:f(r)}\\
V(r)&=&2\mu_1\Phi(p_1r)+2\mu_2\Phi(p_2r)+2\mu_3\Phi(p_3r)+\cdots
\label{eq:V(r)}
\end{eqnarray}
The functions of $f(r)$ and $V(r)$ are the linear combinations of
their lattice-summed functions $\rho(R)$ and $\Phi(R)$, while
the structural dependence is included in the M\"{o}bius
inversion coefficients $\mu_m$ and the radius ratio $p_m$.
Different kinds of functions for $\rho(R)$ and $\Phi(R)$ will be 
used to control the potential convergence, as shown in the next 
section.

\section{Parametrization}

\subsection{$\rho(R)$ and $\Phi(R)$ are exponential functions}

It has been found by Banerjea and Smith using the effective-medium theory
that the off-site electron density exhibits a universal relationship with
respect to lattice spacing: $\rho^{*}=\exp(-a^{*})$, 
which was used to explain the physical origin of the
UBER within the framework of local density approximation~\cite{banerjea-prb}.
Based on the results of Hartree-Fock calculations,
Mei, Davenport and Fernando also pointed out that the lattice sum
of the electron density as a function of lattice constant shows
exponential behaviour.~\cite{mei-prb}
Therefore, it is plausible to take $\rho(R)$ as an exponential

\begin{equation}
\rho(R_1)=\rho_e\exp\left[-\alpha\left(\frac{R_1}{R_{1e}}-1\right)\right]
\label{eq:rhoexp} 
\end{equation}
As a short comment, we would like to point out that when the authors of 
Ref. ~\cite{mei-prb} came to the above equation they just used a complex 
function as $f(r)$=$f_e\sum_{l=0}^k c_l(R_{1e}/r)^l$ to fit it. 
One can see in the present method we do not have to fit. The individual 
function is accurately given by eq.(\ref{eq:f(r)}).

The repulsive energy is often assumed to have a relation with the
bond energy (i.e. the embedding energy in this case) like
$U_{\rm rep}(R)\propto [U_{\rm bond}(R)]^{\gamma}$, where $\gamma$ is
2 according to the so-called Wolfsberg-Helmholtz approximation.
~\cite{pettifor-book}
Therefore, we assume $\Phi(R)$ is also an exponential function

\begin{equation}
\Phi(R_1)=\Phi_e\exp\left[-\beta\left(\frac{R_1}{R_{1e}}-1\right)\right]
\label{eq:phiexp}
\end{equation}
In our method the parameters can be exactly solved as if the model were
a nearest neighbor one. The solutions are

\begin{eqnarray}
\lambda &=&\frac{5GE_s}{3BE_v}\\
&&\nonumber\\
\alpha &=&\sqrt{\frac{\lambda(9\Omega B-15\Omega G)}{E_s-E_v}}\\
&&\nonumber\\
\beta &=&\frac{E_s-E_v}{E_s-\lambda E_v}\alpha\\
&&\nonumber\\
\Phi_e&=&\frac{E_s-\lambda E_v}{\lambda-1}\\
&&\nonumber\\
\rho_e&=&\frac{NS(3)\alpha^3e^{-\alpha}}{8\pi R_{1e}^3}\\
&&\nonumber\\
A&=&\frac{\lambda}{\lambda-1}(E_s-E_v)\rho_e^{-1/\lambda}
\end{eqnarray}

The binding energy equation is a Morse-like function

\begin{eqnarray}
E_{\rm coh}(R_1)=\Phi_e&&\exp\left[-\beta\left(\frac{R_1}{R_{1e}}-1\right)
\right]\nonumber\\
&&-A\rho_e^{1/\lambda}\exp\left[-\frac{\alpha}{\lambda}\left(
\frac{R_1}{R_{1e}}-1\right)\right]
\label{eq:cohexp}
\end{eqnarray}
Different from other equations of state, eq.(\ref{eq:cohexp}) includes the
inputs of the Cauchy pressure and vacancy-formation energy.

\subsection{$\rho(R)$ and $\Phi(R)$ are gaussian functions}

It has been shown that in the present method all the parameters are
analytically determined by the input physical properties, which are
only for the equilibrium lattice. The potential convergence depends on
the functional forms we take for $\rho(R)$ and $\Phi(R)$. 
The gaussian function is an alternative choice

\begin{eqnarray}
\rho(R_1)&=&\rho_e\exp\left[-\alpha\left[\left(\frac{R_1}{R_{1e}}\right)^2
-1\right]\right]\\
\Phi(R_1)&=&\Phi_e\exp\left[-\beta\left[\left(\frac{R_1}{R_{1e}}\right)^2
-1\right]\right]
\end{eqnarray}
The solutions are the same with the above subsection except

\begin{eqnarray}
\alpha&=&\frac{1}{2}\sqrt{\frac{\lambda(9\Omega B-15\Omega G)}{E_s-E_v}}\\
&&\nonumber\\
\rho_e&=&\left(\frac{\alpha}{\pi R_{1e}^2}\right)^{3/2}NS(3)e^{-\alpha}
\end{eqnarray}

The binding energy equation is

\begin{eqnarray}
E_{\rm coh}(R_1)=\Phi_e&&\exp
\left[-\beta\left[\left(\frac{R_1}{R_{1e}}\right)^2-1\right]
\right]\nonumber\\
&&-A\rho_e^{1/\lambda}\exp\left[-\frac{\alpha}{\lambda}\left[\left(
\frac{R_1}{R_{1e}}\right)^2-1\right]\right]
\end{eqnarray}

\subsection{$\rho(R)$ and $\Phi(R)$ are modified exponential functions}

The electron density may not be a simple exponential function. As we
know it is always a combination of some Slater orbitals $r^ne^{-\kappa r}$.
In order to reflect this, we suppose

\begin{equation}
\rho(R_1)=\rho_e\left(\frac{R_1}{R_{1e}}\right)^n
\exp\left[-\alpha\left(\frac{R_1}{R_{1e}}-1\right)\right]
\end{equation}
The pair potential remains the same as eq.(\ref{eq:phiexp}).
The solutions of $\lambda$ are

\[\lambda=\frac{1}{2}\left(1+\lambda_0+n\frac{E_s-E_v}{9B\Omega}\right)\]
\begin{equation}
\pm\frac{1}{2}\left[\left(1+\lambda_0+n\frac{E_s-E_v}{9B\Omega}\right)^2
-4\lambda_0\left(1+n\frac{E_s-E_v}{15G\Omega}\right)\right]^{1/2}
\end{equation}
where $\lambda_0={5GE_s}/{3BE_v}$. In this case, we have two solutions.
Each of them can exactly reproduce the physical inputs. This simple
example then implies that {\em there may exist several different attractors 
leading to different results} when the conventional fitting procedure
is used to search for an approximate solution. This problem may merit a thorough
investigation, and will not be discussed in the present paper.
Note that $(E_s-E_v)/18B\Omega\ll 1$ and $(E_s-E_v)/15G\Omega\ll 1$, 
if $n$ is taken to be 1, then the
approximate solutions will be $\lambda^{+}=\lambda_0$ and $\lambda^{-}=1$.
The latter solution is just the PPM which is then excluded. The solutions for 
the remaining parameters are

\begin{eqnarray}
\alpha&=&n+\sqrt{\frac{\lambda(9\Omega B-15\Omega G)}{E_s-E_v}}\\
&&\nonumber\\
\beta&=&\frac{E_s-E_v}{E_s-\lambda E_v}(\alpha-n)\\
&&\nonumber\\
\rho_e&=&\frac{NS(3)\alpha^{n+3}e^{-\alpha}}{4\pi R_{1e}^3\Gamma(n+3)}
\end{eqnarray}
The expressions for the other two parameter $\Phi_e$ and $A$ are identical
with those presented in the first subsection.

The binding energy equation is

\begin{eqnarray}
E_{\rm coh}(R_1)&=&\Phi_e\exp\left[-\beta\left(\frac{R_1}{R_{1e}}-1\right)
\right]\nonumber\\
&-&A\rho_e^{1/\lambda}\left(\frac{R_1}{R_{1e}}\right)^{n/\lambda}
\exp\left[-\frac{\alpha}{\lambda}\left(
\frac{R_1}{R_{1e}}-1\right)\right]
\end{eqnarray}
When $\alpha/\lambda=\beta$ and $\lambda=n$, the above equation is
just the UBER

\begin{eqnarray}
E_{\rm coh}(R_1)=\frac{\Phi_e}{1-\beta}&&
\left[1+\beta\left(\frac{R_1}{R_{1e}}-1\right)\right]\nonumber\\
&&\times\exp\left[-\beta\left(\frac{R_1}{R_{1e}}-1\right)\right]
\end{eqnarray}

From the above subsections, one can find two points to
support the power-law embedding function. The first
point is that this embedding function (given by $A\rho_e^{1/\lambda}$
and $\lambda$) is independent on the given funcional forms of the 
electron density and pair potential. This physically 
underpins the local nature of the embedding function. It is also true   
when the pair potential and electron density take the power-law forms. 
The second is that the given binding energy equation 
is very close (and even identical) to the UBER. 
This consistency is necessary for a good description for the
thermal expansion (the anharmonity effect)~\cite{foiles-prb88}.
In some sense, the method may represent {\em an embedded-atom 
explanation} for the UBER.

If $\rho(R_1)$ and $\Phi(R_1)$ take the form of power law  
the solutions for the parameters will remain unchanged except that
$\rho_e$ cannot be determined since the power-law function cannot be
normalized. However, an alternative parameter $\xi_e$=
$A\rho_e^{1/\lambda}$ can be determined, as is the case of 
Sutton-Chen's potential. The inverted functions, for example the pair potential,
is given as $V(r)=\Phi_e/S(\beta)(r/R_{1e})^{-\beta}$. While in the case
of exponential, $V(r)<(\Phi_e/12)e^{\beta}[\exp(r/R_{1e})]^{-\beta}$.  
Since $S(\beta)$ is only a little greater than 12 and the paramters
$\Phi_e,\beta$ are the same in the two cases, the above two functions
cross approximately at the NN distance. Beyond the NN distance, the
exponential potential is smaller and decreases much 
faster than the power law.

\begin{figure}
\centerline{\psfig{figure=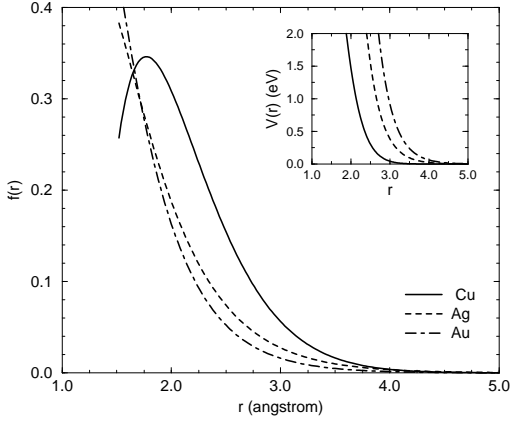,height=6.5cm,angle=0}}
\caption{The inverted electron densities and pair potentials (inset) for the
noble metals. The unit of $f(r)$ is \AA$^{-3}$. }
\label{fig:potential shape}
\end{figure}

Fig. \ref{fig:potential shape} shows the typical shapes of the 
inverted pair potential and electron density, respectively.
One can see from the figure that the inverted electron densities
and pair potentials decrease rapidly.  
In our calculation, the cut-off is placed at about 3$a_e$, where the pair 
potential and electron density have been negligible.

\section{Applications of the potentials}

The physical inputs for the present model are listed in 
Tab. \ref{tab:inputs}. The elastic moduli of Al are from Simmons and 
Wang~\cite{simmons-wang}, those of Ni, Pd, Pt, Cu, Ag and Au are from
Foiles, Baskes and Daws~\cite{foiles-prb86}; Lattice constants and cohesive
energies are all from Kittel~\cite{kittel-book}; Vacancy-formation energies
for fcc transition metals are from Foiles, Baskes and Daw~\cite{foiles-prb86},
and that for Al is from Ballufi~\cite{ballufi}. We do not list
the number of electrons since $\rho_e$ can be incorparated with $A$ 
as a parameter it is not used in monoatomic calculations.
It is only important in alloy calculations, in which it descibes the charge
transfer effect. 

\begin{table}
\caption{
The model inputs $a_{\rm e}$,$E_{\rm s}$,$E_{\rm v}$, $B$,$G$. $a_e$ is in
\AA, $E_s$ and $E_v$ are in eV, $B$ and $G$ are in 10$^{11}$N/m$^2$.}
\begin{tabular}{cccccc}
 Element & $a_e$     & $E_s$ & $E_v$ & $B$ & $G$
\\ \hline
Ni   & 3.52  & 4.44 & 1.7  & 1.804 & 0.95  \\ 
Pd   & 3.89  & 3.89 & 1.54 & 1.95  & 0.54  \\ 
Pt   & 3.92  & 5.84 & 1.6  & 2.83  & 0.65  \\ 
Cu   & 3.61  & 3.49 & 1.3  & 1.38  & 0.55  \\ 
Ag   & 4.09  & 2.95 & 1.1  & 1.04  & 0.34  \\ 
Au   & 4.08  & 3.81 & 0.9  & 1.67  & 0.52  \\ 
Al   & 4.05  & 3.39 & 0.7  & 0.76  & 0.266 \\ 
\end{tabular}
\label{tab:inputs}
\end{table}

\subsection{Structural stabilities}

Phase stability is the first test for the long-range potentials.
For copper, the EAM result $E_{\rm fcc-bcc}$ (25.8meV) falls in the middle
of the nonrelativistic and semirelativistic ab initio values (-17.7 meV 
and -48.8 meV) reported by Lu, Wei and Zunger~\cite{zunger}.
The EAM result for $E_{\rm fcc-diamond}$ equals 1.07 eV,
also close to their ab initio result (1.35 eV) for diamond-like copper
with the correction of nonspherical charge-density inside the muffin-tin
sphere. For all the studied elements, the present model predicts the 
fcc structure to be the ground state (see Tab. \ref{tab:phase stabilities}). 
However, similiar to Sutton and Chen's potentials~\cite{sutton},
the EAM result for $E_{\rm fcc-hcp}$ is virtually zero, so we did not 
print the binding energy curve for hcp in Fig.\ref{fig:phase stability}. 
This failure is believed to be due to the absence of the angularly-dependent 
or higher order moment contributions. It has been pointed
out by Ducastelle and Cyrot-Lackmann that it is mainly the third and fourth 
moments that are responsible for the SEDs among bcc, fcc and hcp 
candidates.~\cite{ducastelle}

\begin{figure}
\centerline{\psfig{figure=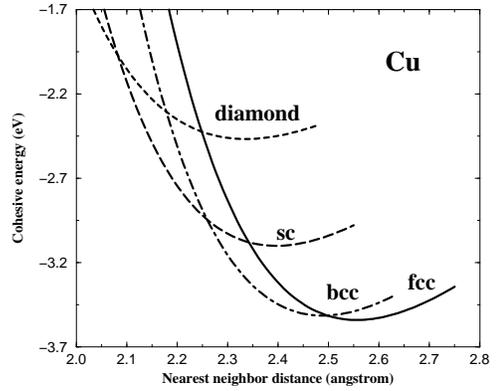,height=6cm,angle=0}}
\caption{The predicted phase stability of copper using the present model.}
\label{fig:phase stability}
\end{figure}

\begin{table}
\caption{The predicted structural energy differences for the fcc
metals. The energies are in eV. The numbers in the parentheses denote
the function types used: 1=exponential; 2=gaussian; 
3=modified exponential.}
\begin{tabular}{cccc}
 Element & $E_{\rm fcc}-E_{\rm sc}$ & $E_{\rm fcc}-E_{\rm bcc}$ &
       $E_{\rm fcc}-E_{\rm diamond}$ \\ \hline
Ni(2) &-0.58 &-5.31$\times 10^{-2}$ & -1.36 \\ 
Pd(1) &-0.53 &-3.42$\times 10^{-2}$ & -1.21 \\ 
Pt(1) &-0.62 &-4.67$\times 10^{-2}$ & -1.39 \\ 
Cu(1) &-0.44 &-2.58$\times 10^{-2}$ & -1.07 \\ 
Ag(1) &-0.40 &-2.28$\times 10^{-2}$ & -0.94 \\ 
Au(1) &-0.35 &-2.29$\times 10^{-2}$ & -0.83 \\ 
Al(3) &-0.28 &-2.11$\times 10^{-2}$ & -0.66 \\ 
\end{tabular}
\label{tab:phase stabilities}
\end{table}

\subsection{Elastic constants and phonon eigenfrequencies}

The comparison of the calculated and experimental data for the
elastic constants $C_{11}$, $C_{12}$, $C_{44}$, the anisotropy ratios
$C/C'$ ($C=C_{44}$, $C'=(C_{11}-C_{12})/2$), and the phonon longitudinal
and transverse frequencies $\gamma_L$, $\gamma_T$ at the boundary of the 
Brillouin zone are shown in Fig. \ref{fig:elastic constants}.
The elastic constants were calculated by exerting the corresponding 
strain matrices to the lattice. The vibrational eigenfrequencies 
are calculated by diagonalizing the EAM dynamical matrix~\cite{daw-ssc}.
The experimental data for the elastic constants of Ni, Pd, Pt, Cu, Ag
and Au are from Foiles, Baskes and Daw~\cite{foiles-prb86}, those for
Al are from Simmons and Wang~\cite{simmons-wang}. The experimental
data for the phonon eigenfrequencies are from Ref.~\cite{adams-jmr}.
The calculated results are overall in agreement with the experimental data.
Fig. \ref{fig:phonon dispersions} shows the predicted phonon dispersion
curves for copper along the high symmetry directions are in excellent 
agreement with the experimental data.

\begin{figure}
\centerline{\psfig{figure=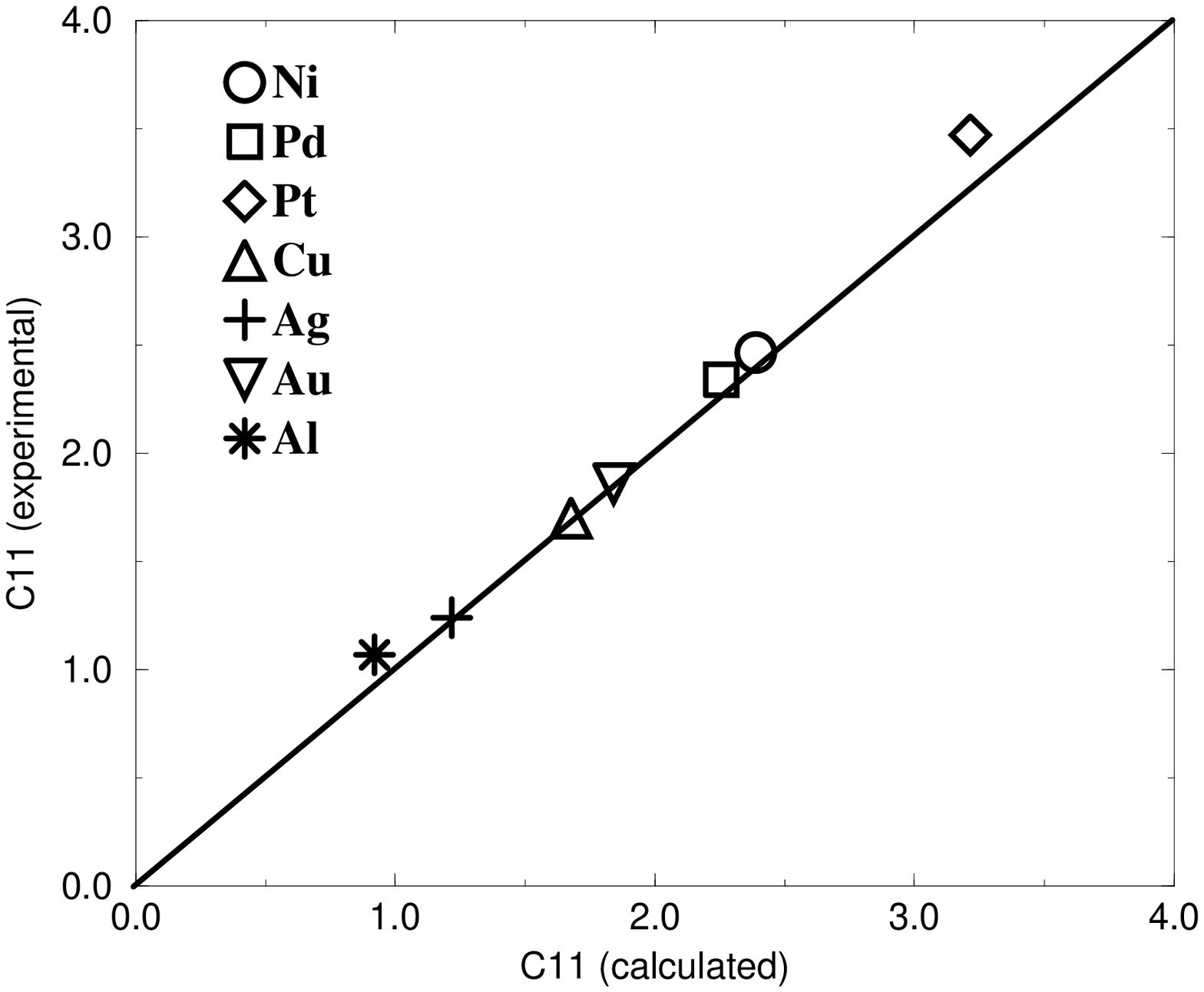,height=3.5cm,angle=0}
\psfig{figure=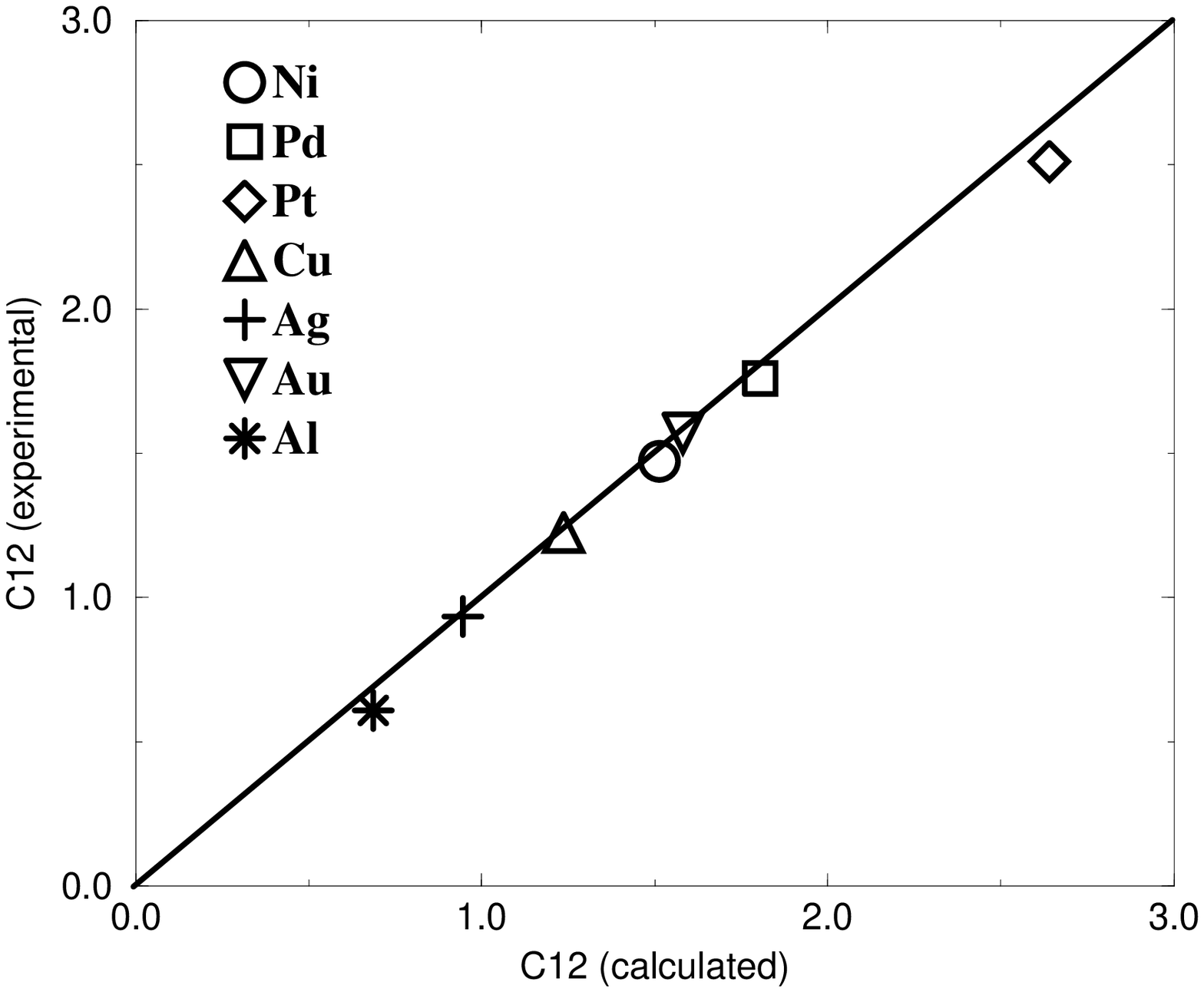,height=3.5cm,angle=0}}
\centerline{\psfig{figure=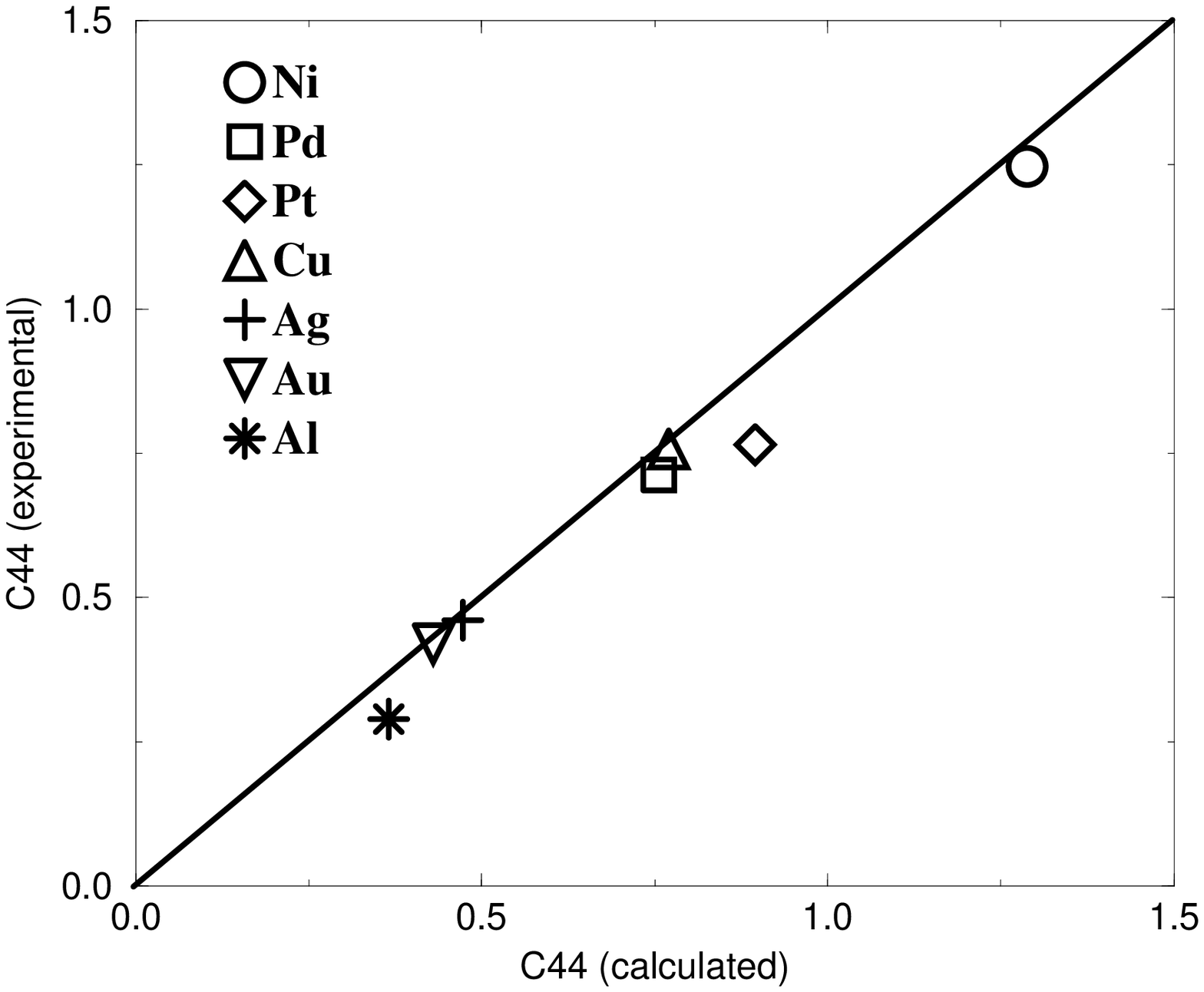,height=3.5cm,angle=0}}
\centerline{\psfig{figure=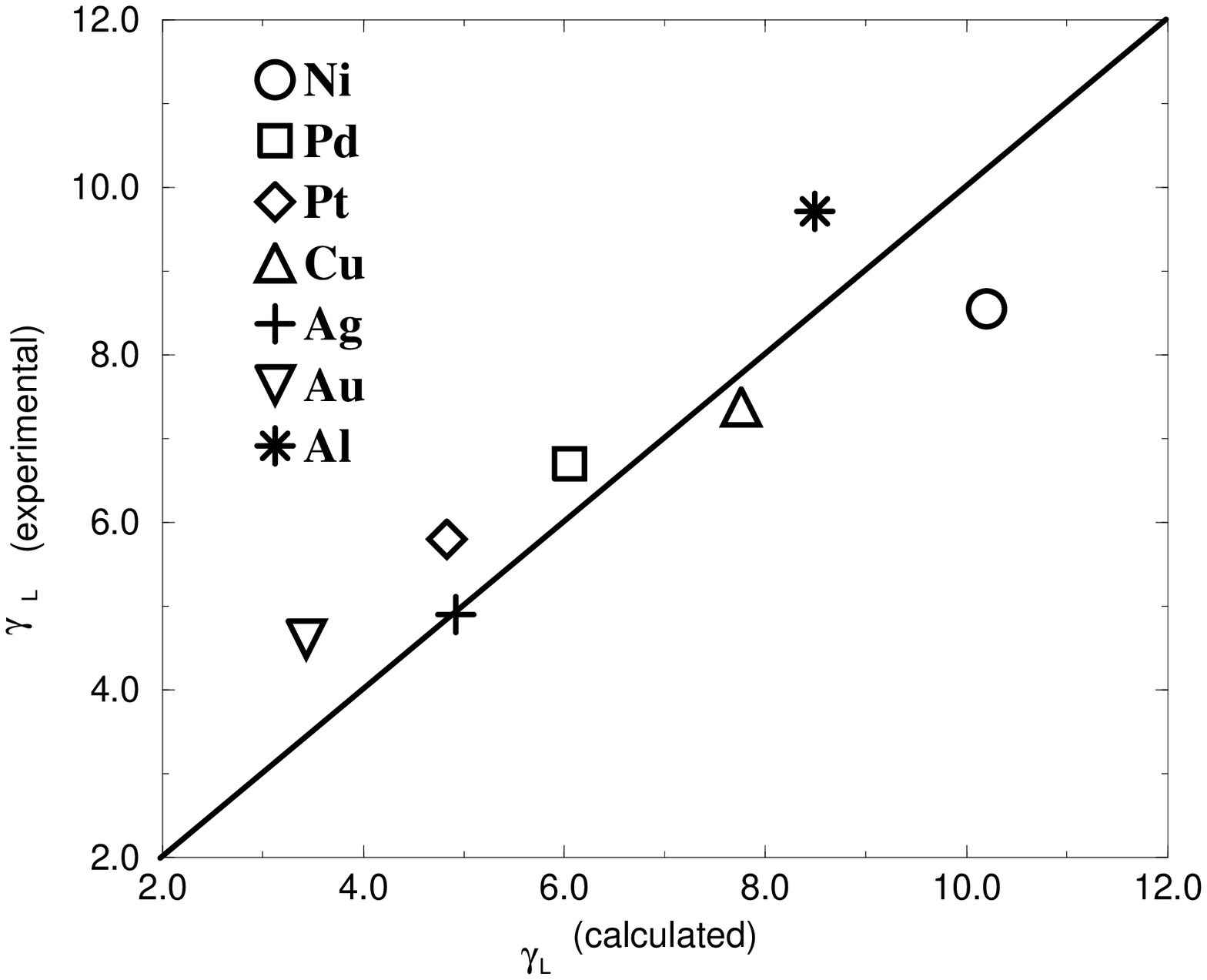,height=3.5cm,angle=0}
\psfig{figure=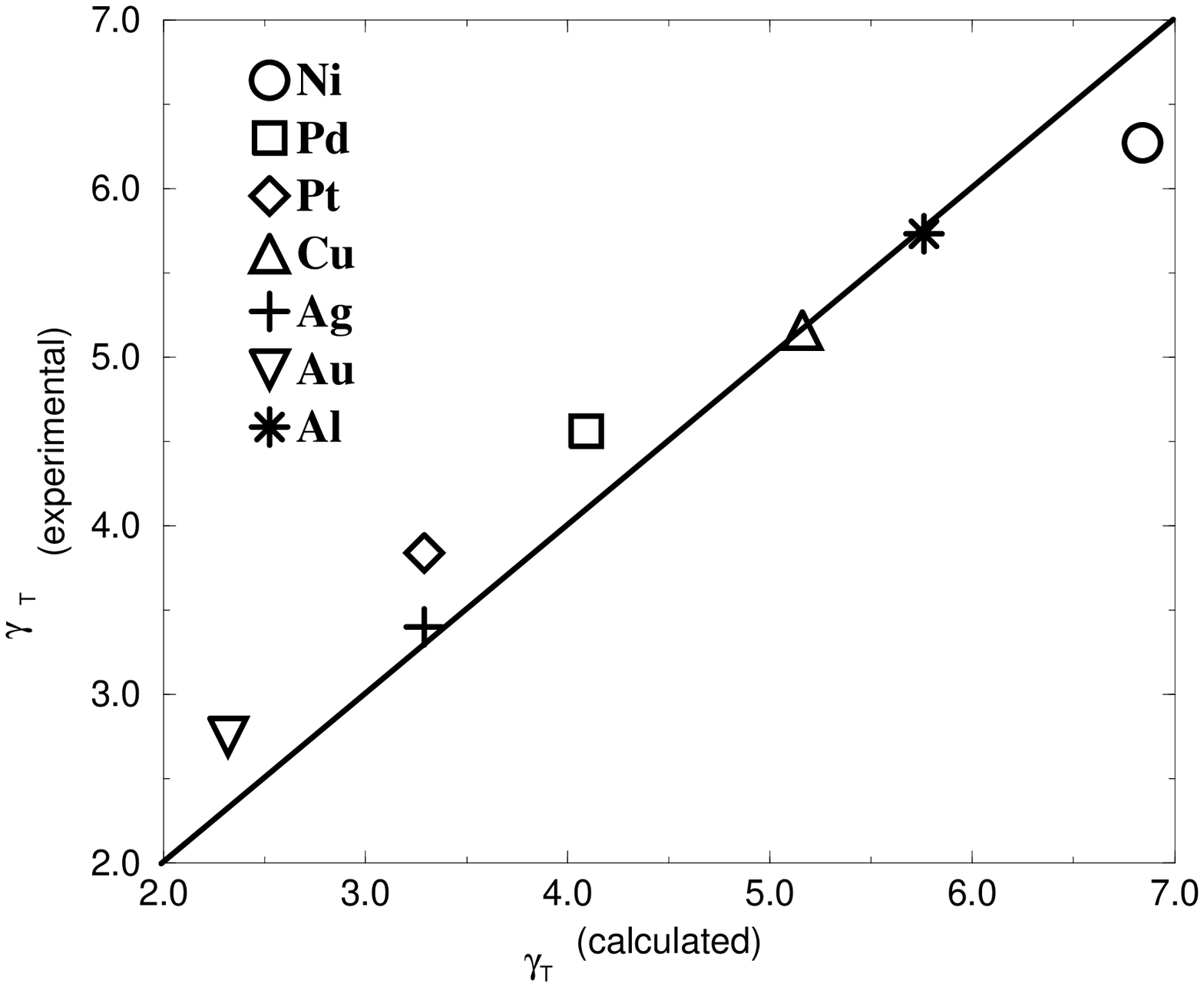,height=3.5cm,angle=0}}
\caption{Comparison of calculated and experimental results for
the elastic constants, the longitudinal and transverse phonon eigenfrequencies
at the Brillouin-zone boundary. (a) $C_{11}$; (b) $C_{12}$; (c) $C_{44}$; 
(d)$\gamma_L$; (e)$\gamma_T$.}
\label{fig:elastic constants}
\end{figure}

\begin{figure}
\centerline{\psfig{figure=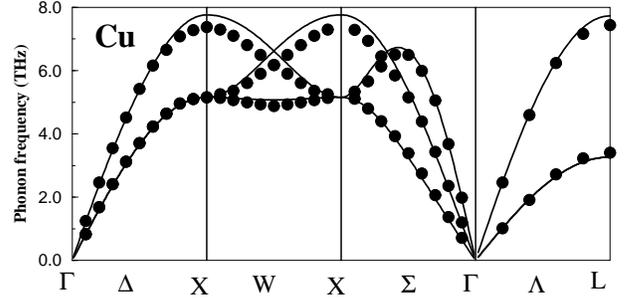,height=7cm,angle=0}}
\caption{Comparison of the theoretical phonon dispersion curves 
of Cu (solid lines) with the experimental data (filled circles) along
the high symmetry directions.}
\label{fig:phonon dispersions}
\end{figure}

It has to be pointed out that when the present model is applied to the bcc
transition metals with low anisotropic ratios, 
the calculated elastic constants $C'$ and
$C_{44}$ severely disagree with the experimental data, despite of that
the bulk modulus and the Voigt shear modulus can be reproduced. This may imply
that for the bcc transition metals the directional bonding is significant.

\subsection{Surface properties}

To calculate surface properties, we employ a simulation box with
size $10\times 10\times 10$ and periodically reproduced in $x$,$y$ and 
half $z$ directions (rather than a slab). For (111) surface, 
a periodic boundary condition with rhombic geometry has been applied.
Relaxation is not considered in the calculation.

The calculated results for the unrelax surface energies of low index
surfaces (100), (110) and (111) are listed in 
Tab. \ref{tab:surface energies}, in comparison with the relaxed results 
of Foiles et al.~\cite{foiles-prb86}. The calculated results for the
adsorption energies $E_{\rm ad}$ at different sites 
(see Fig. \ref{fig:adatom}) and the hopping diffusion barriers 
$U_{\rm diff}$ as well as the island formation energies on the (100) 
surface are given in Tab. \ref{tab:adatom}. The result of $U_{\rm diff}$
for Ag is close to the ab initio results 0.52 eV (LDA) and 0.45 eV (GGA) 
reported by Yu and Scheffler\cite{bdyu}. The binding energies $E_{\rm bind}$
of adatom dimers are calculated to be negative, suggesting that
adatoms tend to form islands. For trimers on the (100) surface, 
the calculated results disfavor the one dimensional configuration.

\begin{figure}
\centerline{\psfig{figure=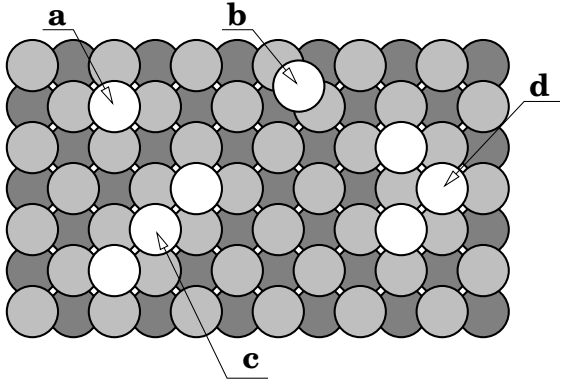,height=5cm,angle=0}}
\caption{Top view of adatoms geometries at (100) surface. (a)
The four-fold hollow site (FFHS); (b) The two-fold bridge site (TFBS); 
(c) One-dimensional (1D) trimer; (d) Two-dimensional (2D) trimer.}
\label{fig:adatom}
\end{figure}

\begin{table}
\caption{The calculated surface energies of low index surfaces. 
The first row are the present results. The second row are the 
theoretical results of Foiles et al.. The experimental data 
(average-face values) as well as the theoretical results 
are all taken from Ref. 27. The units are erg/cm$^2$. }
\begin{tabular}{cccccccc}
 Surfaces & Ni & Pd & Pt & Cu & Ag & Au & Al
\\ \hline
$\gamma_{100}$&1702&1318&1485&1411&782 &891 &614  \\
              &1580&1370&1650&1280&705 &918 &-----\\ 
$\gamma_{110}$&1856&1451&1650&1533&855 &945 &680  \\
              &1730&1490&1750&1400&770 &980 &-----\\ 
$\gamma_{111}$&1595&1181&1286&1320&714 &768 &550  \\
              &1450&1220&1440&1170&620 &790 &-----\\ 
exp.          &2380&2000&2490&1790&1240&1500&----- \\
\end{tabular}
\label{tab:surface energies}
\end{table}

\begin{table}
\caption{The calculated adatom adsorption and island formation
properties on (100) surface. The units are eV.}
\begin{tabular}{llllllll}
 Properties & Ni & Pd & Pt & Cu & Ag & Au & Al
\\ \hline
$E_{\rm ad}$ to FFHS     &-3.59&-3.02&-4.61&-2.77&-2.20&-3.14&-2.85\\ 
$E_{\rm ad}$ to TFBS     &-2.52&-2.48&-4.10&-2.21&-1.76&-2.93&-2.49\\
$U_{\rm diff}$           & 1.07& 0.54& 0.51& 0.56& 0.44& 0.21& 0.36\\ 
$E_{\rm bind}$(dimer)    &-0.42&-0.50&-0.71&-0.37&-0.35&-0.49&-0.29\\
$E_{\rm bind}$(1D trimer)&-0.82&-0.97&-1.35&-0.71&-0.67&-0.94&-0.55\\
$E_{\rm bind}$(2D trimer)&-0.90&-1.02&-1.39&-0.78&-0.72&-0.97&-0.58\\
\end{tabular}
\label{tab:adatom}
\end{table}

\subsection{Molecular dynamics}

In the above subsections the calculations are static. In this subsection,
the potentials are tested in the constant-volume-temperature molecular dynamics
(NVT-MD) simulation for melting processes for Cu and Pt. The simulation box
contains 500 atoms. Gear predictor-corrector algorithm and Verlet
neighbor list are applied. The time step is one 
femtosecond (10$^{-15}$ s). The
ensemble average of the
origin-independent translational order parameter is calculated after
equilibration of at least 5000 steps

\begin{equation}
\sigma^2 = \left\langle\left[\frac{1}{N}\sum_i 
\cos\left({\bf K}\cdot{\bf R}_i\right)\right]^2+
\left[\frac{1}{N}\sum_i\sin\left({\bf K}\cdot{\bf R}_i\right)\right]^2
\right\rangle
\end{equation} 
where $N$ is the number of atoms in the box, ${\bf K}$ is the reciprocal
basis vector for the initial structure (for example, 
${\bf K}=2\pi/a(-1,1,-1)$ for fcc lattice), and ${\bf R}_i$ are the
position vectors for the atoms. 
Fig. \ref{fig:melting} shows the order paramters as a function of temperature
for Cu and Pt. In comparison with experimental data, the melting points are 
underestimated by an amount of 200-300 K. Our result of Cu is worse than 
that of Foiles and Adams~\cite{foiles-adams} (1340 K), but that of Pt is
better than theirs (1480 K). Fig.\ref{fig:g(r)} shows the pair distribution
functions $g(r)$ of Cu at different temperatures.

\begin{figure}
\centerline{\psfig{figure=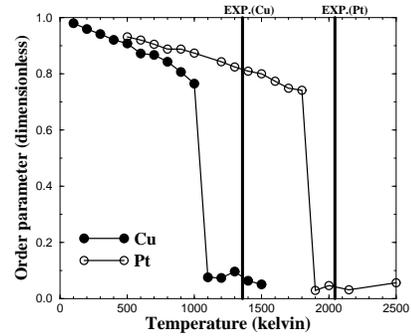,height=5cm,angle=0}}
\caption{Translational order parameters versus temperature for Cu and Pt.
The experimental melting points for Cu(1358K) and Pt(2045K) are denoted
by the two vertical solid lines.}
\label{fig:melting}
\end{figure}

\begin{figure}
\centerline{\psfig{figure=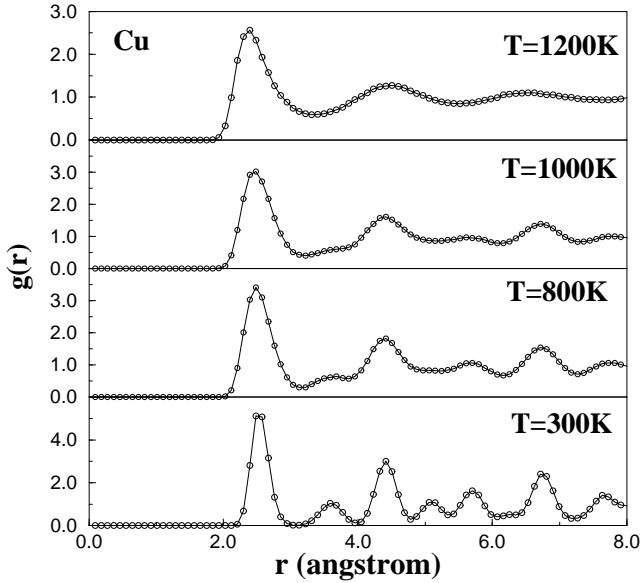,height=8cm,angle=0}}
\caption{Pair distribution functions of Cu at different temperatures.}
\label{fig:g(r)}
\end{figure}

We also simulated the melting of a slab. The size of the simulation box  
is 5$\times$5$\times$20, while that of the slab is 5$\times$5$\times$10
(containing 21 atomic layers or 1050 atoms). The slab is placed at the center of the
simulation box, which is large enough to ensure that the slab does not 
interact with its images. The atomic configuration is described by the
density profile $N(z)$ along the direction perpendicular to the slab. 
$N(z)$ is obtained by averaging over 1000 steps after running
20000 steps. Surface premelting is observed at 900 K, and the liquid
fronts propagate inward when the temperature rises (920K). At 950K
the slab completely melts. 

\begin{figure}
\centerline{\psfig{figure=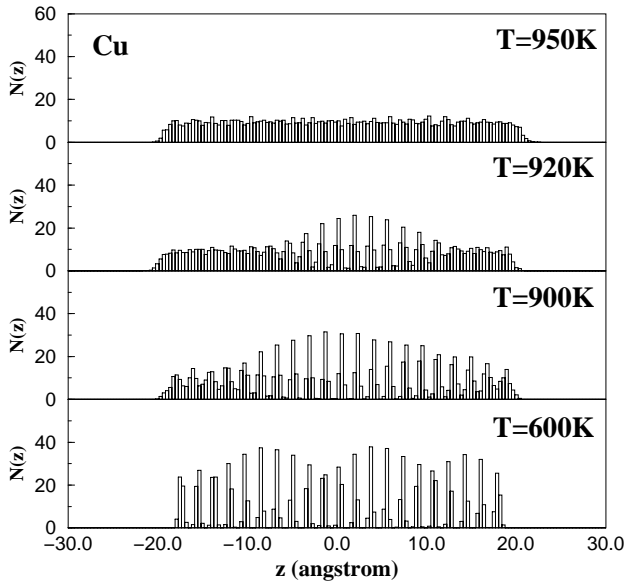,height=8cm,angle=0}}
\caption{Density profiles along the direction perpendicular to the
slab of Cu at different temperatures.}
\label{fig:g(z)}
\end{figure}

The simulation results suggest that melting is a surface-initiated 
process. It would be interesting to investigate the temperature 
dependence of the depth of the melten layer. On the other hand, does 
the solidification process also begin from the surface, or from the
core?   

\section{Concluding remarks}

We have presented a systematical method of obtaining long-range 
embedded-atom potentials. The model parameters 
can be obtained explicitly from six physical inputs and the individual potentials
are inverted from the analytical functions of lattice sums thus arbitary fitting 
can be avoided.  It is shown that the model is able to produce satisfactory 
results of elastic constants, phonon eigenfrequencies, phase stabilities,
surface properties and melting points for the fcc transition metals. The 
potentials are suitable
for computer simulation because of their rapid convergence. 

Deriving interatomic potentials from ab initio calculations when the 
experimental data are not available has become
an abvious trend in the world of material simulation.~\cite{ackland-pma} 
The reason has been explained
very well in a recent paper by Payne et al.~\cite{payne-pmb}. 
In this regard, the present method (as well as the method of Bazant and 
Kaxiras\cite{kaxiras}) may represent an idea 
of bridging the gap between material theory and electronic structure 
theory by the method of 
inverting ab initio EAM potentials (or angularly dependent many-body 
potentials) from first-principles calculations. 
The ab initio binding energy curve can be decomposed
into repulsive and attractive parts, representing the contributions of
the pair potential and embedding energy respectively. 
By using our method, the ab initio EAM potentials can be obtained by 
inverting from the corresponding parts. 

The present model is easy to be generalized to the alloy case by assuming
that the pair potential between unlike atoms is given by Johnson's formula
$V_{ab}(r)=({1}/{2})\{[{f_b(r)}/{f_a(r)}]V_{aa}(r)+[{f_a(r)}/{f_b(r)}]V_{bb}(r)\}$.
~\cite{johnson-prb2} Finally it should be pointed out that the present model 
fails in predicting the bcc transition metals. Modifying the functional
forms of $\rho$, $\Phi$ and $F(\rho)$ does not help much to solve this difficulty.

Acknowledgments: Q.X. gratefully thanks the Max-Planck-Institut f\"{u}r Physik
komplexer Systeme for hospitality. This paper is dedicated to the 60th
Birthday of Prof. N.X. Chen.

\end{document}